\documentclass[doublecol, balance]{epl2} 

\usepackage{epstopdf,graphics}
\usepackage{bm, amsmath}
\usepackage{multirow}
\usepackage{longtable}
\usepackage{balance}
\usepackage{lastpage}

\title{Deep anharmonicity to relativistic spin-0 particles in the spherical regime}
\shorttitle{Title} 

\author{Sami Ortakaya$^{1,2}$}
\shortauthor{S. Ortakaya}
\shorttitle{Relativistic spin-0 particles in the spherical regime}

\institute{                    
  $^{1}$444 Alaska Avenue Suite $\#$BKF475
Torrance, CA 90503 United States\\
$^{2}$Ercis Central Post Office 65400 Van, Turkey\\
{\footnotesize sami.ortakaya@yahoo.com}
}

\abstract{We present an oscillator modeling of the relativistic spin-0 charges moving in the quantum states with minimum coupling of electromagnetic fields. Rather than perturbative approach to spinless regime, we put into operation directly under integer dependent levels for anharmonicity. As a direct solution including a functional procedure to empirical interaction, we consider particle and antiparticle process via quantum numbers. In this way, the charged particle of rest mass energy kept as 280 MeV. Within the familiar Pekeris-like approximation, we have also improved the deep approximation to the orders of $\rm3^{rd}$ and $\rm4^{th}$ near equilibrium of 7.5 fm. Moreover, we have founded a closer agreement of high order approximation and given potential which has width range of 0.43 $\mathrm {fm^{-1}}$. Although equality between scalar and vector potentials give output in the solvable form, the improved approximation provides the spatial-independent rest mass as a \lq\lq{}pure oscillator\rq\rq{} without external field. In the absence of scalar distribution, minimal coupling might also leads to an oscillation at equilibrium distances, so we have considered an adding of extra-energy giving shifted Morse potential in the depth range 80 to 100 MeV. As a result of the shift, it has been concluded that the potential depth of the charged particle affects the relativistic energy levels where we have found about 200 MeV being for particles and nearly -10 MeV being for anti-particles. At the same time, the charge distributions show that the model is also valid for interactions at subatomic distances. Besides negative energy states, the typical probability picture showing spin-zero charge distribution has been followed by the wavefunctions as ($n$=0, $\ell$=0) and ($n$=1, $\ell$=1) corresponding to relativistic energies. By taking into account a deep approximation to Klein-Gordon anharmonicity with $V_{v}(r)$$\neq$0 and $V_{s}(r)$=0, one can introduced approximate-solvable relativistic oscillatory model.}

\begin{document}

\maketitle
\section{Introduction}
In the framework of a stopping material, it can be concluded that negative charges finally moves around an equilibrium. Spatial wave equation of spin-0 scheme have been executed via pi-mesons which have equivalently transformation in the measurable energies. Especially, there is X-ray study of pionic system through particle accelerators \cite{Camac1952}. After considerable evidence of the system, results and collections related to transition energies has been summarized by Backenstoss \cite{Backenstoss1970}. In the following process, both scalar interaction and spin symmetries has been examined in spin-0 and spinor notations, respectively, in which interaction physics is valid, and obtained analytically resolvable forms \cite{Jia2009, Castro2019, Berkdemir2006}. Currently, nonrelativistic molecular physics allows spectral data to be explained \cite{Beuc2018, Jia2015, Sun2013}. In addition to the empirical and solvable forms related to molecular vibrations, the polynomial picture for the subatomic lines might also becomes a permanent point of view.

When charged-particles move in the oscillatory space, atomic interest is also valid for anharmonic wobbling. Rather than the well-known Coulomb interactions, we focus on an empirical form that can also take into account the zero-point energy at femtometer distances.
To analyze existing of the oscillatory modes, one can see that induced electrical dipoles occur in atomic layers. In the case of dipole oscillations, potential minima exist on the quantum well formation. The detailed discussion related to noble gases including dipole oscillations is given in Ref \cite{Kittel1953}.

Without considering molecular polarization, we are directly concerned with the potential field generated by any source towards the charged particle, so we examine relativistic particles under spherical regime of the shifted Morse oscillator. As will seen, we only get radial distribution near equilibrium in the given approach. Primarily, familiar spatial view is that eigenvalue-form is to give a polynomial basis; so that, $n$-quantum number is also obtained for spatial placement \cite{Fleischer1984}. In the presence of the quantum-number dependencies, iterative approach can also accomplish the functional usage with polynomial network \cite{Ciftci2003}. Moreover, algorithmic treatment \cite{Mielnik2000} and symmetry transformation \cite{Fernandez2004} construct solvable models. Regarding integral transform of the given space, integer context have also been introduced within terminal value \cite{Ortakaya2013a} and binomial forms \cite{Chen2004} through space transform. In recent study, we get a functional integral transform into relativistic $q$-deformed anharmonicity including a scalar distribution \cite{Ortakaya2013b}. 

In this study, spin-0 oscillatory regime has been considered in the spherical shell construction. Even though \lq\lq{}scalar equalities\rq\rq{} exhibit solvable wave equations near equilibrium, we can also apply fm-distance to pure quantum states through $V_{s}(r)=0$ and study charged rest mass problem which oscillating in the Morse potential energy by $V_{v}(r)$$\neq$0.

\section{Klein-Gordon Equation Revisited}
We construct a model for the relativistic system including forces which act on the charged particle into consideration. In order to setup an interaction model, one can simplify the particle motion within minimal substitution. Assuming that the induced dipoles exhibit empirical Morse-type oscillatory, a source is considered by a positive charge. Also, negative spin-0 charges expose to forces modeled by oscillator lines, so we focus on he relativistic particle through central Morse oscillator. Here, we consider the Klein-Gordon equation for a charged particle of rest mass $m_{0}$, so $n$\rq{}th eigenvalue spectra can be obtained from \cite{Greiner}
\begin{equation}
\frac{1}{c^2}\left(\mathrm{i}\hbar\frac{\partial}{\partial t}-eA_{0}\right)^2 \Psi_{n}(\vec{\bm{r}})=\left[\left(\bm{p}-\frac{e}{c}\bm{A}\right)^2+m_{0}^2 c^2\right]\Psi_{n}(\vec{\bm{r}}),
\end{equation}
where $e$, $c$ and $\hbar$ are the elementary charge, speed of light and Planck\rq{}s constant, respectively, $\bm{p}$ is the momentum operator defined as covariant derivative $-\mathrm{i}\hbar\bm{\nabla}$ and $A_{\mu}$ denotes the structural four-vector potential $A_{\mu}$=$(A_0,\,0,\,0,\,0)$. We also consider that $eA_{0}=V_{v}(r)$ defines anharmonicity and choose there is no magnetic field: $\bm{A}$=0. In this way, rest mass energy might has a centrifugal dependence given in the form $m_{0}c^2\to m_{0}c^2+V_{s}(r)$. If we only take that $V_{s}(r)$=0, Klein-Gordon equation reads
\begin{equation}\label{e2}
\bigg(-\hbar^2 c^2 \bm{\nabla}^2 + m_{0}^2 c^4\bigg) \Psi(\vec{\bm{r}})=\bigg(E_{n}-V_{v}(r)\bigg)^2\Psi(\vec{\bm{r}}).
\end{equation}
In practice, we refer to radial eigenfunctions derived from $\Psi_{n\ell m}(\vec{\bm{r}})= u_{n\ell}(r)Y_{\ell m}(\theta,\,\varphi)$ with $ u_{n\ell}(r)=\frac{\chi_{n\ell}(r)}{r}$, so we rewrite radial part of Eq.(\ref{e2})
\begin{eqnarray}\label{e2}
\chi_{n\ell}''(r)&+\frac{1}{\hbar^2 c^2}\bigg[\bigg(E_{n\ell}-V_{v}(r)\bigg)^2- m_{0}^2 c^4\nonumber\\&-\frac{\hbar^2 c^2 \ell(\ell+1)}{r^2}\bigg]\chi_{n\ell}(r)=0
\end{eqnarray}
Here, $\ell$=0, 1, 2, 3, $\dots$ is the well-known orbital quantum number on the given space.
\section{Deep Approximation to High-Order}
\begin{figure*}[!hbt]\center
\scalebox{1}{\includegraphics{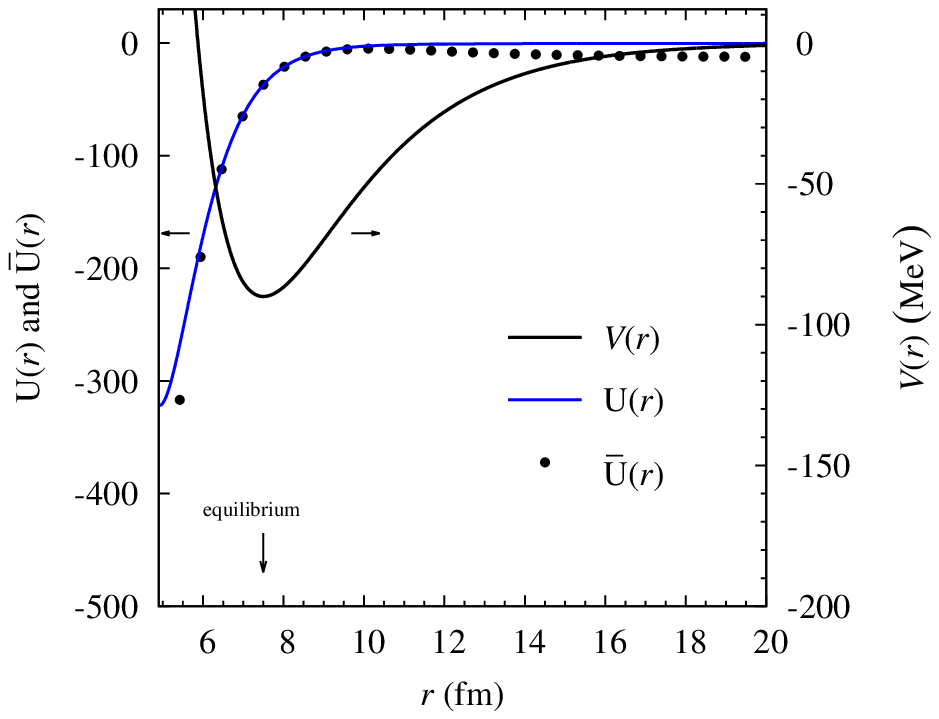}}
\caption{Behavior of the effective potential and approximate form near equilibrium.}\label{f1}
\end{figure*}

The anharmonic motion can be defined via Morse-like energy function in the shifted form
\begin{equation}\label{pot}
V_{v}(r)\equiv V(r)=D_{0}\left(1-\mathrm{e}^{-a(r-r_{e})}\right)^2 -D_{0},
\end{equation}
where $D_{0}$ is the depth of \lq\lq{}quantum well\rq\rq{}, $a$ is the width parameter for spring constant and $r_{e}$ is the equilibrium distance which leads to $V(r_{e})$=$-D_0$. Putting empirical form in Eq. (\ref{pot}) into Eq. (\ref{e2}) and using new variables $x=\frac{r-r_e}{r_e}$ and $\rho=\mathrm{e}^{-a{r_e} x}$, we get
\begin{equation}\label{maine}
\rho^2 \chi\rq{}\rq{}+\rho\chi\rq{}+\left[-\beta_{0}^2+\beta_{1}\rho-\beta_{2}^2 \rho^2 +\frac{\mathrm{U}(r)}{a^2 r_{e}^2}\right]\chi=0,
\end{equation}
where
\begin{eqnarray}
\begin{aligned}
-\beta_{0}^2&=\frac{E^2-m_{0}^2 c^4}{a^2 \hbar^2 c^2},&&\\
\beta_{1}&=\frac{4ED_{0}}{a^2 \hbar^2 c^2},&&\\
-\beta_{2}^2&=\frac{4D_{0}^2-2ED_{0}}{a^2 \hbar^2 c^2}
\end{aligned} 
\end{eqnarray}
and
\begin{equation}
\mathrm{U}(r)=D\rho^4-4D\rho^3-\frac{ \ell(\ell+1)}{(x+1)^2},\qquad D=\frac{D_{0}^2 r_{e}^2}{\hbar^2 c^2}.
\end{equation} 
In order to solve Eq. (\ref{maine}) analytically, we shall use Pekeris-like approximation to spatial dependence of $\mathrm{U}(r)$. For this purpose, we can use that the high-order terms approach to
\begin{equation}
{\rm{U}}(r)\simeq\overline{\rm U}(r)=A_{0}+A_{1}\rho+A_{2}\rho^2 .
\end{equation}
as a deep anharmonicity. ${\rm{U}}(r)$ and $\overline{\rm U}(r)$ can be expanded as series of $r$ near $r=r_{e}\,(x=0)$. Comparing them up to third-order, we obtain
\begin{gather}
\begin{aligned}
A_0 &= \frac{3\gamma\alpha-\gamma\alpha^2-3\gamma}{\alpha^2}-D, &&\\
A_1 &= \frac{6\gamma-4\gamma\alpha}{\alpha^2}+4D, &&\\
A_2 &= \frac{\gamma\alpha-3\gamma}{\alpha^2}-6D,
\end{aligned}
\end{gather}
where $\gamma$$=$$\ell(\ell+1)$ and $\alpha=ar_{e}$. ${\rm U}(r)$ and $\overline{\rm U}(r)$ can be compared when $D_{0}$=$90\,{\rm MeV}$ at $r_{e}$=$7.5\, {\rm fm}$. We can see from Fig. \ref{f1} that, besides ${\rm U}(r)$ with inverse square $1/r^2$, $\overline{\rm U}(r)$ is also valid near $r$=$r_{e}$. The deep formation yields a closeness for width range $a r_{e}$=3.2 with spring-constant parameter, $a$=$0.43\, \rm{fm^{-1}}$.

Inserting approximate part into Eq. (\ref{maine}), we rewrite the solvable equation
\begin{equation}\label{maine2}
\rho^2 \overline{\chi}\rq{}\rq{}+\rho\overline{\chi}\rq{}+
\bigg[-\overline{\beta}_{0}^2+\overline{\beta}_{1}\rho-\overline{\beta}_{2}^2 \rho^2\bigg]\overline{\chi}=0,
\end{equation}
where
\begin{eqnarray}
\begin{aligned}
-\overline{\beta}_{0}^2&=\frac{E^2-m_{0}^2 c^4}{a^2 \hbar^2 c^2}+\frac{A_0}{a^2 r_{e}^2},&&\\
\overline{\beta}_{1}&=\frac{4ED_{0}}{a^2 \hbar^2 c^2}+\frac{A_1}{a^2 r_{e}^2},&&\\
-\overline{\beta}_{2}^2&=\frac{4D_{0}^2-2ED_{0}}{a^2 \hbar^2 c^2}+\frac{A_2}{a^2 r_{e}^2}.
\end{aligned}
\end{eqnarray}
If we let $\overline{\chi}(\rho)$ provided to be
\begin{equation}
\overline{\chi}(\rho)=\rho^{-\overline{\beta}_{0}}f(\rho);\;\;\overline{\beta}_{0}>0
\end{equation}
then Eq. (\ref{maine2}) gives
\begin{equation}\label{maine3}
\rho f\rq{}\rq{}(\rho)+(1-2\overline{\beta}_{0})f\rq{}(\rho)+
\bigg[\overline{\beta}_{1}-\overline{\beta}_{2}^2 \rho\bigg]f(\rho)=0.
\end{equation}
Note that $\overline{\chi}(\rho)$ approaches zero as $\rho\to0$ $(r\to\infty)$, so we must get
\begin{equation}
f(\rho)\propto\rho^{\sigma};\;\;\sigma>\overline{\beta}_{0}.
\end{equation}
As can be seen from transformed-space \cite{Ortakaya2013a}, Eq. (\ref{maine3}) gives $n$\rq{}th eigenvalues given by
\begin{equation}\label{nn}
\frac{\overline{\beta}_{1}}{2\overline{\beta}_{2}}-\overline{\beta}_{0}=n+\frac{1}{2},\qquad(n=0,\,1,\,2\,\dots)
\end{equation}
and we should have Kummer\rq{}s solution 
\begin{equation}
\overline{\chi}(\rho)=\rho^{\overline{\beta}_{0}}{\rm e}^{-\overline{\beta}_{2}\rho}\Phi\left(-n,\,2\overline{\beta}_{0}+1;\,2\overline{\beta}_{2}\rho\right).
\end{equation} 
\section{Numerical Results}
\begin{figure*}
\centering
\scalebox{0.85}{\includegraphics{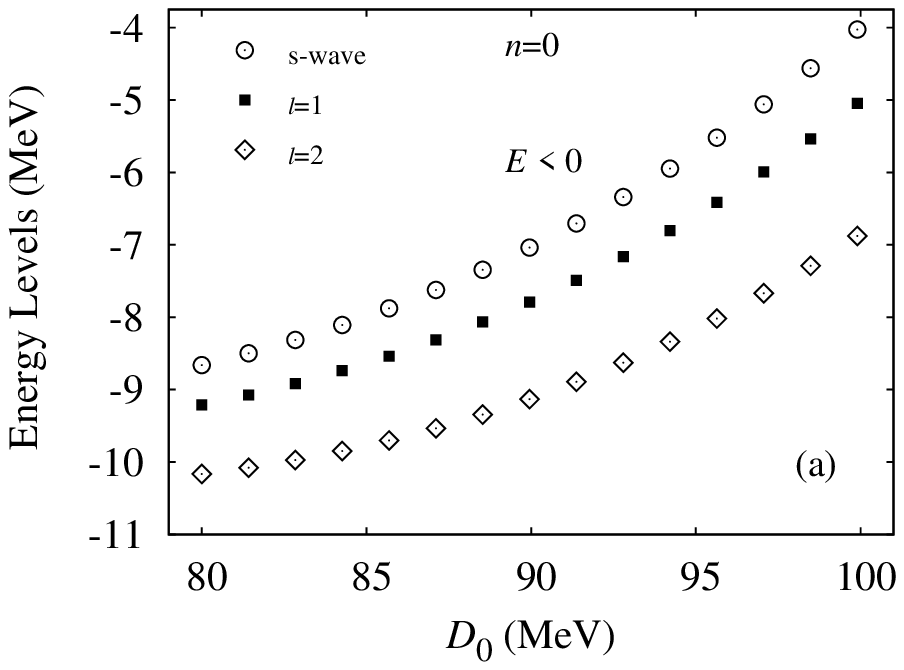}}\scalebox{0.85}{\includegraphics{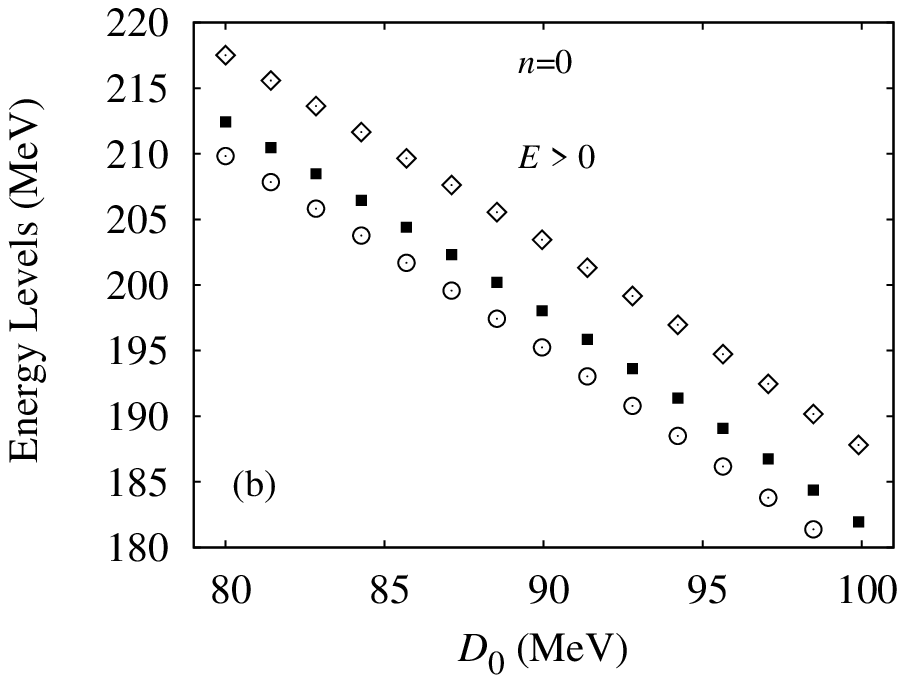}}\\
\scalebox{0.85}{\includegraphics{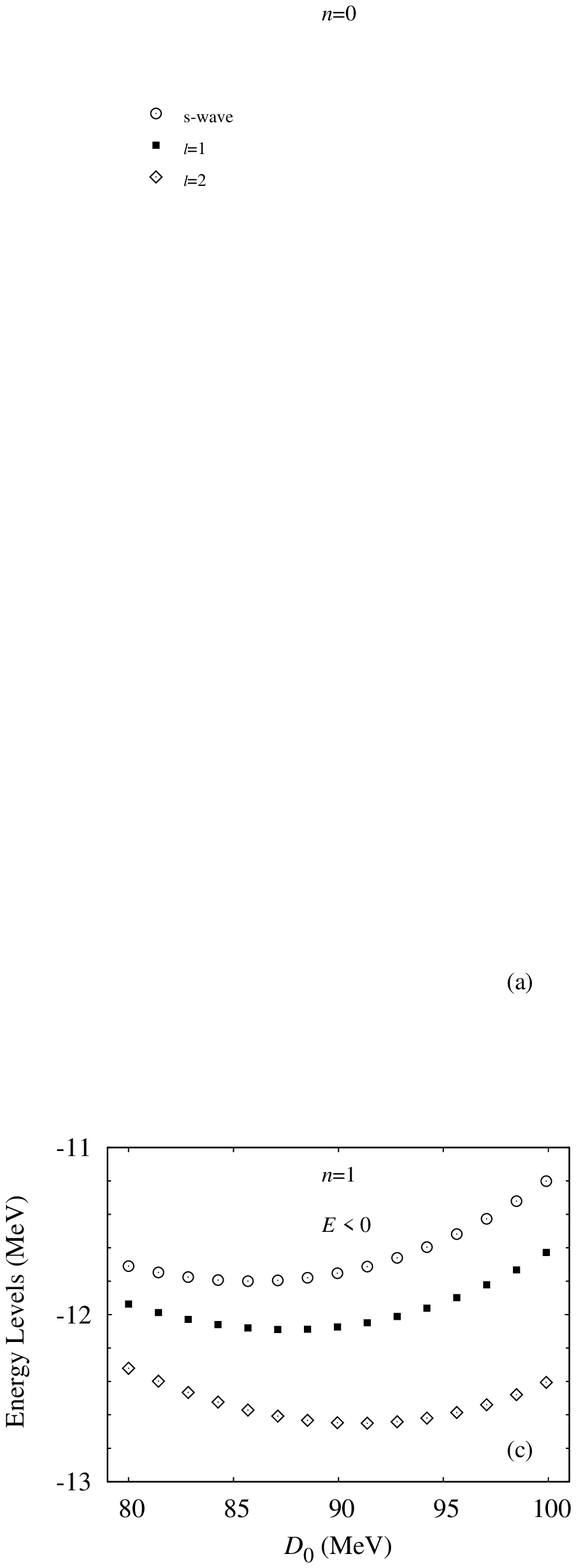}}
\scalebox{0.85}{\includegraphics{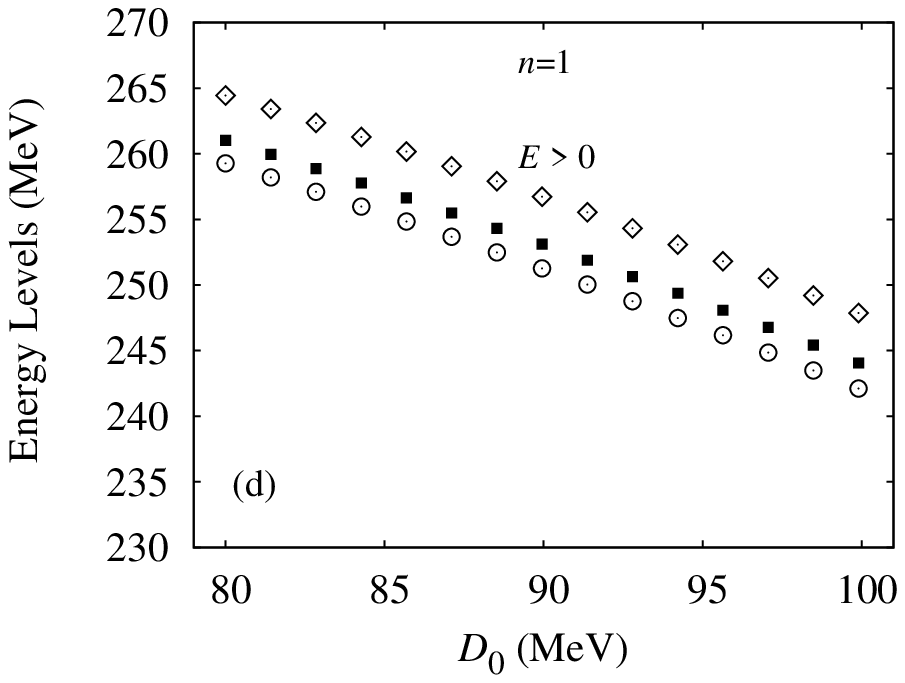}}
\caption{Relativistic energy levels for the Morse-type anharmonicity when rest mass is 0.30075211 u which corresponds to 280 MeV.}\label{d0}
\end{figure*}

\balance
Solving the transcendental equation given in Eq. (\ref{nn}), it can be seen that energy levels depend on the depth of anharmonicity. Fig. \ref{d0} shows eigenvalues as a function of depth $D_{0}$ at MeV-scale of equilibrium distance 7.5 fm. Assuming width range is ${\rm 0.43\, fm^{-1}}\,\,(ar_e=3.2)$, relativistic energies lie the ranges $$180\, {\rm MeV}<E_{n\ell}<270\, {\rm MeV}$$
$$ {\rm and}$$
$$ -13\,{\rm MeV}<E_{n\ell}<-4\,\rm MeV$$
for particle and antiparticle, respectively. The obtained dependencies are consistent with spherical symmetry related to atomic numbers \cite{Fleischer1984}. Also, excited energies shift to large values for particles and lead to small energies for antiparticles. It can be concluded that MeV-scale eigenvalues satisfy the range $-m_{0}c^2$$<$$E_{n\ell}$$<$$+m_{0}c^2$ when $V_{s}$$=$0. Rather than critical values, increasing (or decreasing) of the excited states shows that antiparticle and particle states are expected as $E$$<$0 and $E$$>$0, respectively. Then, anharmonicity related to high-energies can be easily provided to be effective. 

Within the effective potential form, varying radially mass taken as $m_{0}c^2$$\to$$m_{0}c^2+V_{s}(r)$, produce a more general usage for equality phenomena. In the case of equal scalar and vector potentials, ground-state $|00\rangle$ lies the energy about 310 MeV (at $r_e$$=$7.5 fm with $D_0$$=$90 MeV). It is also much larger than the constant-mass regime $V_{s}$$=$0. As a result of the constant rest mass, bound states for $|E_{n\ell}|$$<$$m_{0}c^2$ occur when it taken the values of $r_{e}$$=$7.5 fm, $\alpha$$=$3.2 and $D_{0}$$\,\sim\,$90 MeV. In a way, bound states can be expected for the effective potentials has \lq\lq{}well structure\rq\rq{} related to Morse-type anharmonicity (see Ref. \cite{Castro2019}).

\begin{figure*}[!hbt]\center
\scalebox{.8}{\includegraphics{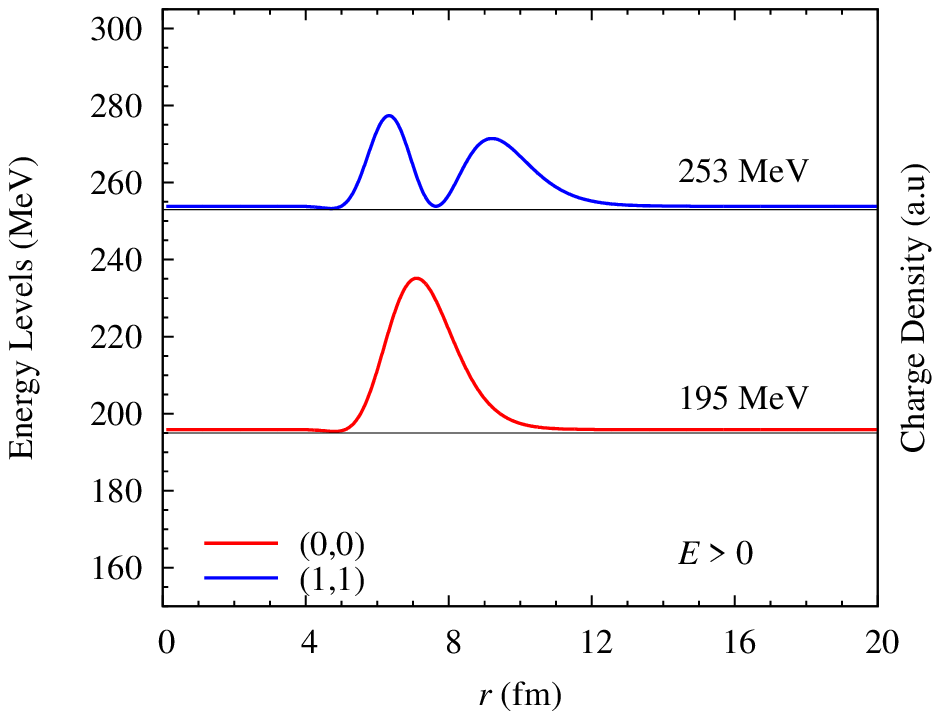}}\scalebox{.8}{\includegraphics{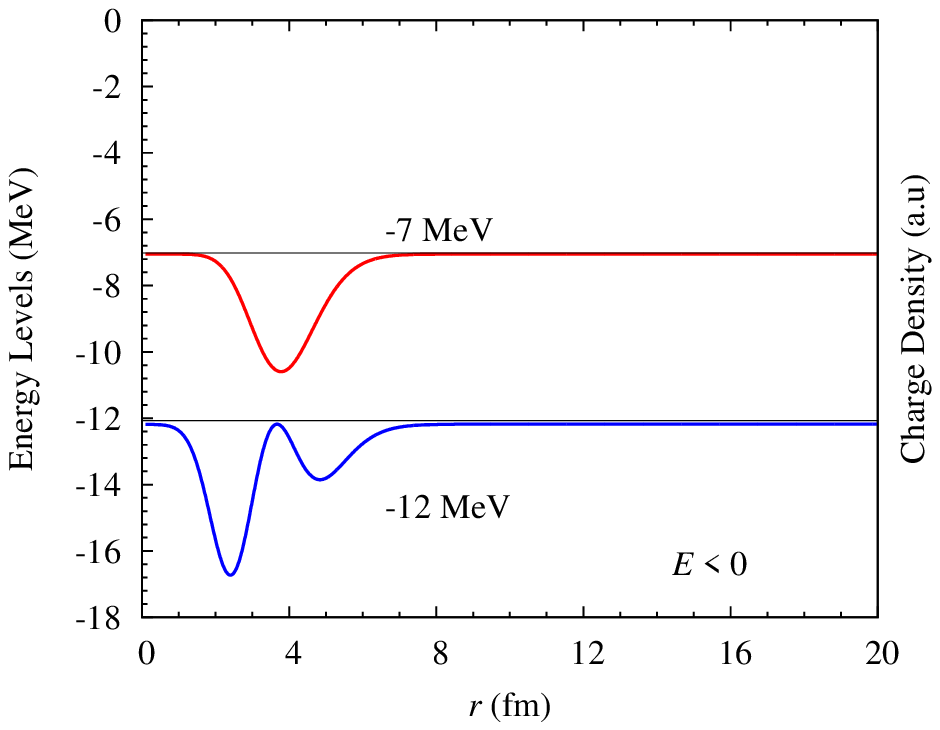}}
\caption{Charge densities for the relativistic spin-0 particles moving in the anharmonic oscillator. Here (0, 0) and (1, 1) denote arbitrary ($n,\,\ell$) states.}\label{wp}
\end{figure*}

Now we get wave-functional behaviors for charge density given as $$\varrho(r)=\pm \frac{e|E|}{mc^2}u^{*}_{\pm}u_{\pm}.$$
Here the solutions which $+$ being for particles and $-$ is being for antiparticles, given in Fig. \ref{wp}. When keeping $D_0$$=$90 MeV at $r_e$$=$7.5 fm and varying $|n\ell\rangle$  as $|00\rangle$ and $|11\rangle$, the densities are localized near $r_e$. Comparing $E$$<$0 and $E$$>$0, a shift to smaller distance at the maxima can be seen when antiparticle ones. Also, a result of the localized behavior is that approximation is valid near $r_e$ and radial distribution leads to zero.

\section{Conclusion}
In the framework of the relativistic Morse oscillator, spin-0 eigenvalues have proved to be a quantum anharmonicity is valid for MeV scale at about 10 fm width. Relating the results ($V_{s}$=0, $V_{v}$$\neq$0) to the equality regime ($V_{s}$$=$$V_{v}$), bound states occur easily considering width value of 0.43 $\rm fm^{-1}$. At fm-scale space, the key feature regarding spin-0 regime is to accomplish the deep approximation to the spherical oscillatory with inverse square (or effective potential) including the orders of $\rm 3^{rd}$ and $\rm 4^{th}$. In conclusion, bound states of relativistic oscillator with constant rest mass are obtained and physical wave function near equilibrium also showed charge density.




\newpage
\balance

\newpage
\end{document}